\documentclass[12pt,tightenlines,eqsecnum,floats,aps,amsmath,amssymb,nofootinbib,prd]{revtex4}

\usepackage{setspace}
\usepackage{subfig}
\usepackage{amsmath,amssymb,amsfonts,amsthm,mathrsfs}
\usepackage{mathtools}
\usepackage{graphicx}
\usepackage{enumerate} 
\usepackage{color}
\usepackage[cal=boondoxo]{mathalfa}

\DeclareCaptionJustification{justified}{\leftskip=0pt \rightskip=0pt \parfillskip=0pt plus 1fil}
\def\e{{\bf e}}                                                       
\def\a{{\bf a}} 
\def\k{{\bf k}}

\begin{document}

\title{Axisymmetric gravity in real Ashtekar variables:\\ the
  quantum theory}

\author{Rodolfo Gambini$^{1}$, Esteban Mato$^{2}$, Jorge Pullin$^{3}$}
\affiliation {
1. Instituto de F\'{\i}sica, Facultad de Ciencias, 
Igu\'a 4225, esq. Mataojo, 11400 Montevideo, Uruguay. \\
2. Instituto de F\'{\i}sica, Facultad de Ingenier\'{\i}a, 
J. Herrera y Reissig 565, 11300 Montevideo, Uruguay.\\
3. Department of Physics and Astronomy, Louisiana State University,
Baton Rouge, LA 70803-4001, USA}

\begin{abstract}
In a previous paper we formulated axisymmetric general re\-la\-ti\-vi\-ty in terms of real
Ashtekar--Barbero 
va\-ria\-bles. Here we proceed to quantize the theory. We are able to
implement Thiemann's version of the Hamiltonian constraint. 
This provides a $2+1$ dimensional arena to test ideas
for the dynamics of quantum gravity and opens the possibility of
quantum studies of rotating black hole spacetimes.
\end{abstract}
\maketitle
\section{Introduction}

In a previous paper \cite{kerr1} we laid out the framework for
Ashtekar--Barbero variables for general relativity with one
axisymmetric Killing vector field. Here we would like to discuss the
quantization of the model. We will implement Thiemann's Hamiltonian
\cite{thiemann} and show that its algebra is consistent. This provides
a framework to discuss other implementations of the Hamiltonian like
those proposed by Varadarajan \cite{varadarajan} or Lewandowski and
collaborators \cite{lewandowski}. It provides the simplest environment
where those types of dynamics can be implemented capturing key
elements of the full three dimensional implementation. Simpler
examples like spherical symmetry or cosmologies do not allow to
implement all the features needed in the full theory. From a
phy\-si\-cal point of view, the axisymmetric case is of great
relevance in astrophysics. Our treatment opens the possibility of
considering the quantum space-times associated with black holes with
angular momentum, for example.

The organization of this paper is as follows. In section 2 we present
a summary of the classical theory. In section 3 we recall some
properties of loop quantization for the full theory. In section 4 
we introduce the kinematics of the axisymmetric case. In section 5 we
discuss the implementation of the Euclidean portion of the Hamiltonian
constraint. In section 6 we analyze the Lorentzian portion. In section
7 we prove cylindrical consistency, diffeomorphism covariance and
freedom from anomalies. We end with conclusions in section 8.

\section{Summary of the classical theory}

Following our previous treatment \cite{kerr1},
we choose adapted spatial coordinates $(x,y,\phi)$ and the Killing
vector is $\partial_\phi$. In this framework one can decompose the
triads and the connections in terms of reduced variables,
\begin{align}
 A&=&A_a^i\tau_i \mathrm{d}x^a= \left((\cos(\phi)\tau_1 + \sin(\phi) \tau_2)  \a_a^1 + (-\sin(\phi) \tau_1 + \cos(\phi) \tau_2) \a_a^2 + \a_a^3 \tau_3 \right)\mathrm{d}x^a \\
 E&=&E^a_i \tau^i \partial_a= \left((\cos(\phi)\tau^1 + \sin(\phi) \tau^2) \e^a_1 + (-\sin(\phi) \tau^1 + \cos(\phi) \tau^2) \e^a_2 + \e^a_3 \tau^3 \right)\partial_a,
\end{align}
where $\tau_i=-\frac{i}{2} \sigma_i$ with $\sigma_i$ the Pauli matrices.

The symmetry adapted variables $(\a_a^i,\e^b_j)$ do not depend
on the angular coordinate $\phi$, i.e. only on $(x,y)$, and are
canonically conjugate, i.e.,
\begin{equation}
\label{poisson_bracket_axi}
\{\a^i_a(\vec x),\e^b_j(\vec x')\}=4 G\beta\,\delta^i_j\delta^b_a\delta^{(2)}(\vec x-\vec x'),
\end{equation}
where $\beta$ is the Immirzi parameter. It should be noted that
although the symmetry adapted variables only depend on $x,y$ their
indices run from 1 to 3.

The total Hamiltonian in terms of the reduced variables is,
\begin{equation}
h_T=\frac{1}{8 G}\left[g(\vec\Lambda)+ d(\vec N)+c(N)\right],
\end{equation}
and the (smeared) constraints are,
\begin{align}
g(\vec\lambda)&=\frac{1}{\beta}\int dxdy \lambda^i\left(\partial_a\e^a_i+\varepsilon_{ijk}\a^j_a\e^a_k+\varepsilon_{ijk}\delta^j_3\delta^\phi_a \e^a_k\right),\\
d(\vec N)&=\frac{1}{\beta}\int dxdy N^a\left(\e^b_i\partial_a\a^i_b-\partial_b(\e^b_i\a^i_a)+\delta_a^\phi\delta^i_3\varepsilon_{ijk}\a^j_b\e^b_k
\right), \\
h_E(N)&=-2\int dxdy \frac{N}{\sqrt{\e}}\left[(\a^{i}_{b,a} - \a^{i}_{a,b} + \epsilon_{ilm}\a^{l}_{a} \a^{m}_{b})\epsilon_{ijk}\e_{j}^{a} \e_{k}^{b} +2\delta^j_3 \delta_b^\phi\left(\a^i_a \e_i^a\e_j^b - \a^i_a\e_i^b \e_j^a\right)\right] ,\\
h_L(N)&=2\int dxdy \frac{N}{\sqrt{\e}} (1+\beta^2)\epsilon_{ijk} \epsilon_{ilm} \e_{j}^{a} \e_{k}^{b} \k^{l}_{a} \k^{m}_{b},
\end{align}
where  we have written the scalar constraint as
$c(N)=h_E(N) + h_L(N)$.

\section{Summary of loop quantization}

We review some general concepts of loop quantum gravity, more details
can be found in \cite{thiemann}
In the full theory one considers densitized triads $E^a_i$ and
SU(2) connections $A_a^i$. One can construct a quantum connection
representation in which the actions of the basic variables are,
\begin{eqnarray}
 A_a^i(x) \psi[A] &=&  A_a^i (x) \psi[A], \\
  E^a_i(x) \psi[A] &=& -i 8\pi G \hbar \beta \frac{\delta}{\delta A_a^i (x)} \psi[A].
\end{eqnarray}
It is useful to consider smeared versions of the variables, for
instance the holonomy of the connection along a closed curve $\gamma$, 
\begin{align}
 H[A,\gamma]=P\left( e^{\int_\gamma A^i_a \tau_i \mathrm{d}x^a} \right),
\end{align}
and the flux of the triad through a surface $S$ parametrized by
$\sigma_1$ and $\sigma_2$,
\begin{align}
\label{flujo_triada}
 E_i= \int_S n_a E^a_i = \int_S \mathrm{d}\sigma_1 \mathrm{d} \sigma_2 \varepsilon_{abc} \frac{\partial x^b (\vec{\sigma})}{\partial \sigma_1} \frac{\partial x^c (\vec{\sigma})}{\partial \sigma_2} E^a_i.
\end{align}
We will need to consider the algebra of these quantities given that
the triad and connection are canonically conjugate,
\begin{align}
 \left\lbrace A_a^i (x) , E^b_j (y) \right\rbrace = 8\pi G \beta \delta_a^b \delta_i^j \delta^3(x-y).
\end{align}
This leads to,
\begin{align}
\label{flujo}
 \left\lbrace E_i (S) , H[A,\gamma] \right\rbrace = \pm 8\pi G \beta H[A,\gamma_{1s}] \tau_i H[A,\gamma_{s0}],
\end{align}
where we introduced the notation $H[A,\gamma_{s_1 s_2}]=P\left(
  e^{\left( \int_{s_1}^{s_2} A_a^i \tau_i \dot{\gamma}^a(t)
      \mathrm{d}t \right)} \right)$ for the parallel transport operator, $\gamma(s)$ is the intersection
point of $S$ and $\gamma$ and $1$, $0$ are the values of the parameter
of the origin of the loop. The expression also holds for open paths,
in that case $1$ would be the origin and $0$ the endpoint. The sign is given by the relative
orientation between them and can be computed through the integral,
\begin{align}
 \int_S \int_\gamma \mathrm{d}\sigma^1 \mathrm{d}\sigma^2 \mathrm{d} s
  \varepsilon_{abc} \frac{\partial x^a (\sigma)}{\partial \sigma^1} \frac{\partial x^b (\sigma)}{\partial \sigma^2} \frac{\partial x^c (s)}{\partial s} \delta^3 (\vec{x}(\vec{\sigma}),\vec{x}(s))=\pm 1.
\end{align}

\section{Kinematical quantum theory for axisymmetry}

We would now like to repeat some of these constructions for the
axisymmetric theory in terms of the reduced variables. As we
mentioned, the reduced variables are ${\bf a}_a^i(x,y)$ and ${\bf
  e}^a_i(x,y)$. With $a=x,y,\phi$. The spatial Killing vector field is perpendicular to a family of two-surfaces which contain the rotation axis. The coordinates $(x,y)$ are defined on one of these surfaces (which we will denote by $\Sigma$) and can be carried to the rest of the spacetime along the integral curves of the Killing vector field (see for example \cite{wald_book}, sec. 7.1). We will
consider holonomies along curves either contained in $\Sigma$ or transverse to it. Also the surfaces $S$ we will
consider can either be contained in $\Sigma$ or are transverse and
of the form
$\gamma_a\times [0,\mu]$ where $\gamma_a$ is a curve contained in
$\Sigma$ and $\mu$ is a real number, an interval in the  $\phi$ coordinate.

We start by considering the Poisson bracket of the flux through one of
these latter surfaces 
with a holonomy along a curve in $\Sigma$,
\begin{align}
 \left\{E_{i}(S), H[A, \gamma]\right\}=\int_{S} \int_{\gamma} \mathrm{d} \sigma^{1} \mathrm{d} \sigma^{2} \mathrm{d} s \varepsilon_{abc} \frac{\partial x^{a}}{\partial \sigma^{1}} \frac{\partial x^{b}}{\partial \sigma^{2}} \frac{\partial x^{c}}{\partial s} \delta^{3}(\vec{x}(\vec{\sigma}), \vec{x}(s))\left[H\left[A, \gamma_{1 s}\right] \tau_{i} H\left[A, \gamma_{s 0}\right]\right].
\end{align}
One of the coordinates that parametrizes the surface (say, $\sigma_2$)
corresponds to the direction transverse to $\Sigma$; we can
therefore identify it with the coordinate $\phi$. Integrating in it
just picks the value of $\phi$ in which the curve and the surface
intersect, which is just the value of $\phi$ corresponding to the two-surface transverse to $\partial_\phi$ that we have chosen
\begin{align}
 \left\{E_{i}(S), H[A, \gamma]\right\}=\int_{\gamma_\alpha} \int_{\gamma} \mathrm{d} \sigma \mathrm{d} s \varepsilon_{\alpha\beta} \frac{\partial x^{\alpha}}{\partial \sigma} \frac{\partial x^{\beta}}{\partial s} \delta^{2}(\vec{x}(\sigma), \vec{x}(s))\left[H\left[A, \gamma_{1 s}\right] \tau_{i} H\left[A, \gamma_{s 0}\right]\right].
\end{align}
One can carry out the integrals more or less like in the general case
and obtain an analogous result.
\begin{align}
\label{flujo_axi}
 \left\{E_{i}(S), H[A, \gamma]\right\}=\pm H\left[A, \gamma_{1 s}\right] \tau_{i} H\left[A, \gamma_{s 0}\right].
\end{align}
(Where we have switched to units such that the factor $4G\beta$ in \eqref{poisson_bracket_axi} equals one).
We will now consider holonomies along curves $\gamma^\phi$ generated by the Killing vector field, 
\begin{align}
 H\left[A, \gamma^{\phi}\right]=e^{\int \mathrm{d} \phi \a_{\phi}^{i}(x, y) \tau_{i}}=e^{ \mu \a_{\phi}^{i}(x, y) \tau_{i}}
\end{align}
where $\mu$ is the length of the curve. These are actually point holonomies since under a gauge transformation $g$: 
\begin{align}
H\left[A, \gamma_{\phi}(x, y)\right] \rightarrow U(g(x, y)) H\left[A, \gamma_{\phi}(x, y)\right] U^{-1}(g(x, y))
\end{align}
with $U(g)$ the gauge transformation's representation. It is easy to
see that the Poisson bracket between one of these holonomies and the
flux on a surface whose normal vector is of the form $n_\alpha\in \Sigma$ vanishes due to the $\varepsilon$ factor.
Let us now consider the case of  a surface contained in the plane and
a curve perpendicular to it of arbitrary origin and length, but such that they intersect at one point. Calling
the intersection point $s$ one has that,
\begin{align}
 \left\{E_{i}(S), H[A, \gamma^\phi]\right\}=\pm H\left[A,
  \gamma^{\phi}_{1s}\right] \tau_{i} H\left[A, \gamma^{\phi}_{
  s0}\right].
\end{align}

Later, in order to build gauge invariant spin networks states we will
need to consider transverse holonomies along complete circles that
start and end in the plane. In that case the previous result reduces
to,
\begin{align}
 \left\{E_{i}(S), H[A, \gamma^\phi]\right\}=\pm H\left[A, \gamma^{\phi}\right] \tau_{i}. 
\end{align}

As is usual in loop quantum gravity we wish to consider kinematical quantum states
that are cylindrical functions of parallel transports of the
connection. Given a series of open paths $\gamma^n$ we have,
\begin{align}
 \Psi_{\Gamma,f}=f\left( H[A,\gamma^1] , \dots , H[A, \gamma^N] \right).
\end{align}

A basis of gauge invariant states for this space is given by spin networks. The latter
can be defined as a triplet $S=(\Gamma,j_l, i_n)$ with $\Gamma$ a
graph with $L$ edges and $N$ nodes.  The quantity $j_l$ indicates the
representation of the parallel transport matrix along
$\gamma_l$. The spin network state has the form,
\begin{align}
 \Psi_S[A]=\left\langle A | S \right\rangle = \left( \bigotimes_{l=1}^L \left( H^{j_l}[A,\gamma^l] \right) \right) \cdot \left( \bigotimes_{n=1}^N i_n \right).
\end{align}
\indent The intertwiners $i_n$ have a set of indices dual to that of the
product of parallel transport matrices and their contraction with them
ensures gauge invariance of the state.

For the symmetry reduced theory we will consider curves
$\gamma^\alpha$ contained in the $x,y$ two-surface and $\gamma^\phi$ curves
perpendicular to it. The spin network states take the form,
\begin{align}
 \Psi_S[A]=\left\langle A | S \right\rangle = \left( \bigotimes_l \left( H^{j_l}[A,\gamma^{\alpha_l}] \right) \right) \cdot \left( \bigotimes_n i_n \right) \cdot \left( \bigotimes_k \left( H^{j_k}[A,\gamma^{\phi_k}] \right) \right).
\end{align}
The parallel transports along the curves $\gamma^\phi$ take the form
of point holonomies, as we discussed,
\begin{align}
\label{grasp_transversal}
H^{j_l}[A,\gamma_\phi]=P\left( e^{\int \mathrm{d}\phi \a_\phi^i (\tau_i)^{j_l}} \right)= e^{2 \pi n \a_\phi^i (\tau_i)^{j_l}},
\end{align}
where ${\bf a}_\phi^i$ is evaluated in a point $x,y$ of the two-surface and
$n$ is a natural number that corresponds to the number of turns of the
curve $\gamma^\phi$ around the axis of symmetry.\\[10pt]
\begin{figure}[h]

{\includegraphics[width=9cm]{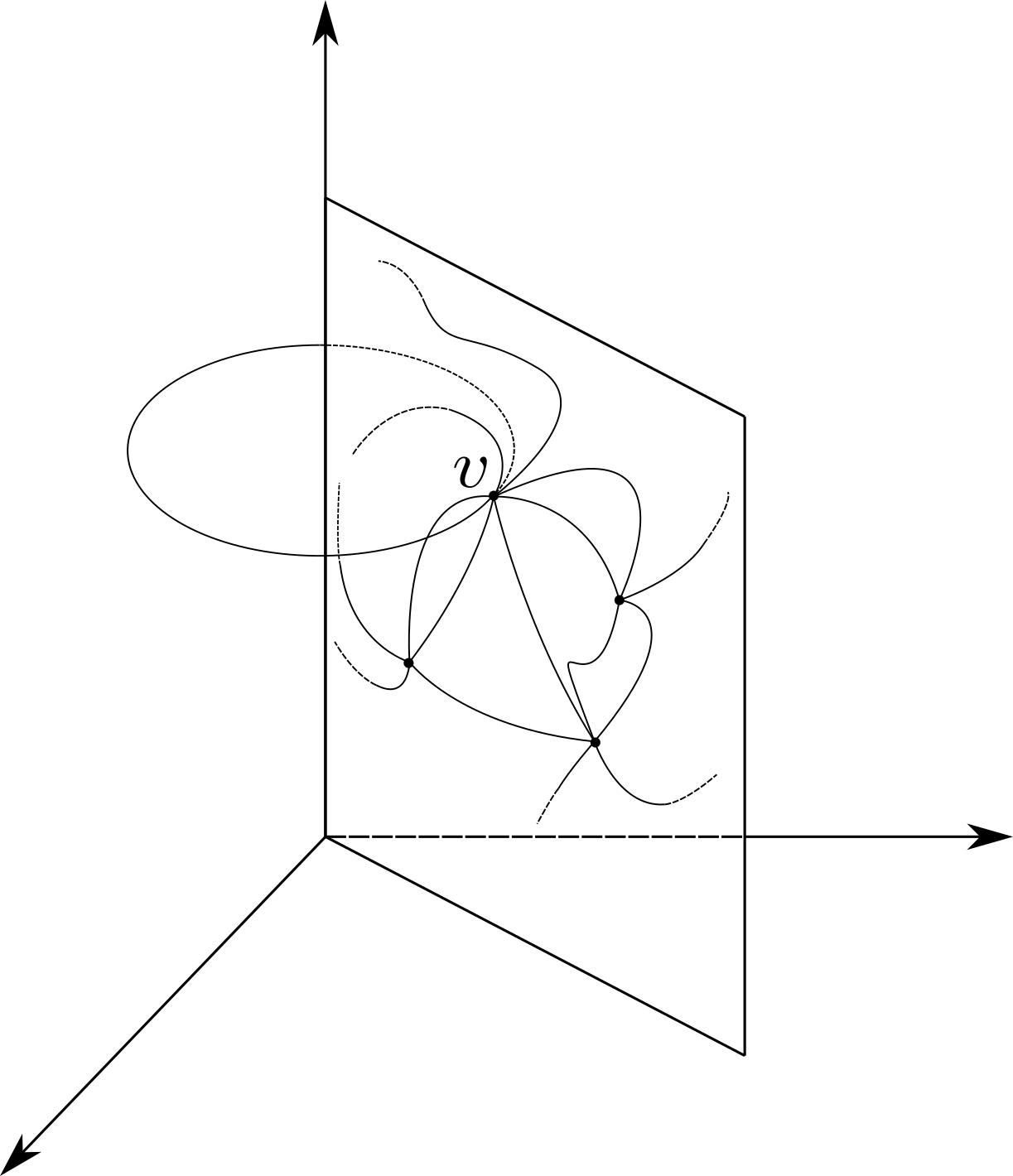}} 

\caption{
Example of a graph on which the spin-networks in the reduced
theory are based. The plane actually represents any two-surface
orthogonal to the Killing vector field (we chose a plane for
simplicity). The edges orthogonal to the plane start and end at the
same vertex.}
\end{figure}

Let us turn our attention to the area operator. The classical
ex\-pre\-ssion for the area of a surface in terms of the triads is,
\begin{align}
 A(S)=\int_S \sqrt{n_a E^a_i n_b E^b_i} \mathrm{d}^2 \sigma.
\end{align}
This can be promoted to an operator in the quantum theory using the
ex\-pre\-ssions we discussed for the action of the flux of the triads on
parallel transport matrices. The end result on a spin network state
based on a graph $\Gamma$ is well known,
\begin{align}
 \hat{A}(S) \psi_\Gamma[A] = \sum\limits_{P\in S\cap \Gamma} \sqrt{j_p (j_p+1)} \psi[A],
\end{align}
with $j_p$ the valence of the edge of $\Gamma$ that intersects the
surface $S$ at the point $P$.

In the reduced theory it only makes sense to consider surfaces
compatible with the symmetry. That is, either surfaces $\Sigma$ perpendicular to the Killing vector field or surfaces of the form $\gamma\times
\Phi$ where $\gamma$ is an edge contained in the two-surface $\Sigma$ and
$\Phi$ is an interval of the real line along $\phi$. It is relatively
straightforward to see that the expression
of the area operator is the same as in the full theory, only with the
spin network living in the two-surface $\Sigma$ with possible transverse
insertions of curves $\gamma^\phi$. For areas lying in the $x,y$ plane one will only have
non-vanishing contributions from the insertions.

For the definition of the Hamiltonian constraint it is important to
have the definition of the volume operator. In the full theory it is
given by \cite{asle_volume},
\begin{align}
 \hat{V} \Psi_\Gamma = \kappa_0 \sum\limits_v \sqrt{\hat{q}_v} \Psi_\Gamma,
\end{align}
where the sum is over all the vertices of the graph $\Gamma$ and $\kappa_0$ is an arbitrary constant. The
operator $\hat{q}_v$ is given by
\begin{align}
 \hat{q}_v  = \frac{1}{48} \varepsilon_{ijk} \sum\limits_{e,e',e''}\epsilon \left(e,e',e''\right) J^i_{v,e} J^j_{v,e'} J^k_{v,e''},
\end{align}
with the $J$'s the angular momentum associated to a point $x$ in the
manifold and an edge $e$ that emerges from it,
 \begin{align}
 \left(J_{x, e}^{i} \Psi_{\Gamma}\right) &=i\left(H[A,\Gamma] \tau^{i}\right)_{B}^{A} \frac{\partial \psi_\Gamma}{\partial\left(H[A,\Gamma]\right)_{B}^{A}} \nonumber \\ &=i \frac{d}{d t}\left[\psi_\Gamma \left( \ldots, H[A,\gamma] \exp \left(t \tau^{i}\right), \ldots \right)\right], \label{momento_angular}
 \end{align}
 with $A,B$ indices in any representation of SU(2).
In the above expression the sum extends to all triplets of edges $e,e',e''$ that intersect in a
vertex. The quantity $\epsilon(e,e',e'')$ is $\pm 1$ if the edges are
linearly independent and zero otherwise. 

The expression of the volume for the axisymmetric case is essentially
the same as in the full theory, except that one of the three edges is
required to be in the transverse direction in order to have a
non-vanishing volume.

The Gauss constraint is automatically satisfied by considering spin
networks, as they are gauge invariant. The diffeomorphism constraint is discussed in the following section. 

\section{Implementation of the diffeomorphism constraint}

When written in terms of our reduced variables (2.1, 2.2), the diffeomorphism constraint takes the following form:
\begin{align}
 d(\vec{N})=\frac{2}{\beta} \int d x d y N^{a}\left(\mathbf{e}_{i}^{b} \partial_{a} \mathbf{a}_{b}^{i}-\partial_{b}\left(\mathbf{e}_{i}^{b} \mathbf{a}_{a}^{i}\right)+\delta_{a}^{\phi} \delta_{3}^{i} \varepsilon_{i j k} \mathbf{a}_{b}^{j} \mathbf{e}_{k}^{b}\right)
\end{align}
and its action on the canonical variables is given by \cite{kerr1}:
\begin{align}
 \left\{\mathbf{a}_{a}^{i}(\vec{x}), d(\vec{N})\right\}=\left.\left(\mathbf{a}_{b}^{i} N_{, a}^{b}+N^{b} \mathbf{a}_{a, b}^{i}+\varepsilon_{i j k} N^{d} \delta_{d}^{\phi} \delta_{3}^{j} \mathbf{a}_{a}^{k}\right)\right|_{\vec{x}}
\end{align}
Let us first analyze the diffeomorphisms in the direction orthogonal
to the Killing vector field, that is, shifts of the form
$ N=N^\alpha \partial_\alpha $, with $\alpha=x,y$ (we will focus on
the action of the constraint on the connection, everything we argue
below applies equally to the triad). The previous action reduces to:
\begin{align}
 \left\{\mathbf{a}_{a}^{i}(\vec{x}), d(N^\alpha \partial_\alpha)\right\}=\left.\left(\mathbf{a}_{\alpha}^{i} N_{, a}^{\alpha}+N^{\alpha} \mathbf{a}_{a, \alpha}^{i}
 \right)\right|_{\vec{x}}
\end{align}
The right hand side is just the Lie derivative of the connection in the direction of $N^\alpha \partial_\alpha$, the action of the diffeomorphism constraint in a direction transverse to the Killing Vector Field on the reduced variables is thus completely analogous to the one on the old variables. \\
The case of the diffeomorphisms in the direction of the Killing vector field, although it is also analogous, is a little more subtle and requires some discussion. 

Let us begin by noting that the action of the constraint
\begin{align}
 \left\{\mathbf{a}_{a}^{i}(\vec{x}), d(N^\phi \partial_\phi)\right\}=\left.\left(\mathbf{a}_{\phi}^{i} N_{, a}^{\phi}+N^{\phi} \underbrace{\mathbf{a}_{a, \phi}^{i}}_{0}+\varepsilon_{i j k} N^{d} \delta_{d}^{\phi} \delta_{3}^{j} \mathbf{a}_{a}^{k}\right)\right|_{\vec{x}}
\end{align}
does not even look like a Lie derivative, so it would appear that it is more complicated than the previous case. However, it is easy to show that after considering the connection $\tilde{\a}_a^i \coloneqq \a_a^i + \delta_a^\phi $ and redefining a multiplier in the Gauss constraint, we obtain
\begin{align}
 \left\{\tilde{\mathbf{a}}_{a}^{i}(\vec{x}), d(N^\phi \partial_\phi)\right\}=\left.\left(\tilde{\mathbf{a}}_{\phi}^{i} N_{, a}^{\phi}+N^{\phi} \underbrace{\tilde{\mathbf{a}}_{a, \phi}^{i}}_{0} \right)\right|_{\vec{x}}
\end{align}
which is just the Lie derivative in the direction of
$N^\phi \partial_\phi$. So in reality the action of the constraint was
just obscured by the use of $\a_a^i$ instead of $\tilde{\a}_a^i$,
which should be expected since the former doesn't even transform as a
connection under the action of the Gauss constraint while the later
does (see \cite{kerr1}). The redefinition of the Lagrange multiplier
corresponding to a Gauss constraint should also be expected since we
arrived at the reduced variables by following a typical reduction
procedure for connections:
\begin{align}
 \mathcal{L}_{\tilde{K}} A_a^i = \epsilon_{i j k} \lambda^{j} A_{a}^{k}
\end{align}
being $\tilde{K}=\lambda \partial_\phi$ and $\vec{\lambda}=(0,0,\lambda)$, with $\lambda$ a constant (the simplest choice possible). The previous equation amounts to
\begin{align}
 \partial_{\phi} A_{a}^{i}=\epsilon_{i 3 k} A_{a}^{k}
\end{align}
so we see that our requirement actually mixes a term corresponding to
the Lie derivative with the action of the Gauss constraint. 

In summary, the action of the diffeomorphism constraint in the
direction of the Killing vector field is analogous to that of the full
theory (with the only exception that there is homogeneity in that
direction so that scalar functions would be left invariant by the
constraint (but not one-forms and vectors). The constraint should
therefore be implemented in the same way as in the full theory, two
2-surfaces which can be deformed into each other by dragging points
along the Killing direction (the finite action of the constraint)
should be considered equivalent.

Imposing the diffeomorphism constraints in the directions
orthogonal to the Killing orbits has an analogue effect to that of the full theory,
namely, we obtain that all graphs which are knotted in the same way in
the two-surface are equivalent to each other.

\section{The Euclidean portion of the Hamiltonian constraint}

The Hamiltonian constraint of general relativity written in terms of
Ashtekar--Barbero variables contains two terms, the Euclidean and
Lorentzian ones,
\begin{align}
C(N) &= H_E(N)+H_L(N).
\end{align}
We will discuss the Lorentzian part later. For the Euclidean portion
Thiemann \cite{thiemann} showed that it can be written as,
\begin{align}
\label{hamiltoniano_reescrito}
 H_E(N) = 2 \int_{\Sigma} \mathrm{d}^3 x N(x) \varepsilon^{abc} \operatorname{tr}\left( F_{ab} \left\lbrace A_c , V \right\rbrace \right).
\end{align}

Thiemann also showed how a discretization of  this classical expression can be written in terms of
holonomies. Let us review his construction in general before
proceeding to the reduced theory. One partitions the manifold in elementary tetrahedra
$\Delta$. In each tetrahedron we choose a vertex and call it
$v(\Delta)$. Let $s_i(\Delta)$ with $i=1,2,3$ the edges that arrive at
$v$ and let $\alpha_{ij}=s_i(\Delta)\circ a_{ij} \circ s^{-1}_j(\Delta)$
a loop based in $v(\Delta)$ with $a_{ij}$ an edge that links the
endpoints of $s_i$ and $s_j$ that do not coincide with $v(\Delta)$. 
One considers the quantity 
\begin{align}
\label{h_triangulacion}
 H_T^E (N) = \sum\limits_{\Delta \in T} H_\Delta^E (N),
\end{align}
with,
\begin{align}
\label{h_triangulacion_2}
 H_\Delta^E (N) \coloneqq -\frac{2}{3} N_v \varepsilon^{ijk} \operatorname{tr} \left( h_{\alpha_{ij}(\Delta)} h_{s_k (\Delta)} \left\lbrace h^{-1}_{s_k (\Delta)} , V \right\rbrace \right).
\end{align}
It can be readily seen that this quantity tends to $2\int_{\Delta} N
{\rm tr}(F\wedge \{A,V\})$ when the tetrahedron $\Delta$ shrinks to
its base point $v(\Delta)$. All of the quantities involved in that
expression (holonomies, volume
operator) can be promoted to quantum operators. Replacing each
quantity by its corresponding operator and the Poisson bracket by
$i\hbar$ the commutator we get,
\begin{align}
 \hat{H}_{T}^{E}(N) :=\sum_{\Delta \in T} \hat{H}_{\Delta}^{E}(N), \hat{H}_{\Delta}^{E}(N) & :=-2 \frac{N(v(\Delta))}{3 i \ell_{p}^{2}} \epsilon^{i j k} \operatorname{tr}\left(h_{\alpha_{i j}(\Delta)} h_{s_{k}(\Delta)}\left[h_{s_{k}(\Delta)}^{-1}, \hat{V}\right]\right) \nonumber \\ & =: N_{v} \hat{H}_{\Delta}^{E},
\end{align}
with $\ell_{p}$ Planck's length. Although we have chosen $4 G\beta=1$
we have kept $\beta$ and $\ell_{p}$ explicit in various places without
replacing their values for clarity.

It can be shown that when applied to a function $\Psi_\Gamma$ cylindrical with respect to the graph $\Gamma$, the Hamiltonian acts only on the vertices of $\Gamma$:
\begin{align}
 \hat{H}^E_T (N) \Psi_\Gamma = \sum\limits_{v\in \Gamma} N_v \hat{H}^E_v \Psi_\Gamma,
 \end{align}
with $ \hat{H}^E_v \coloneqq \sum\limits_{v(\Delta)=v} \hat{H}^E_\Delta $. This quantity is finite as long as the number of tetrahedra that
adjoin the vertices of the graph $\Gamma$ is maintained finite when one
refines the partition. There are many possibilities when choosing a partition. One possible prescription is the following: given an edge
$e_I$ of $\Gamma$ with origin at $v$ and outgoing from it, we choose
as an edge of $\Delta$ a segment $s_I$ with an end in $v$  which is
contained in $e_I$, the other end does not coincide with the end of
$e_I$ that is not $v$. Given an unordered pair of edges $s_I$ and
$s_J$ of $\Delta$, let $a_{IJ}$ be a curve that joins the end of those
edges that are not $v$; $a_{IJ}$ does not intersect $\Gamma$ at any other
point. For each unordered triplet of edges $s_I$, $s_J$, $s_K$ that
intersect at $v$ and define a tetrahedron $\Delta$  \cite{thiemann}, one can define in a
diffeomorphism invariant way seven other tetrahedra (called mirror
tetrahedra) that together with $\Delta$ saturate the vertex $v$ \cite{thiemann}. With
this one can complete the regularization of the operator. We define
$D(\Delta)$ a closed region covered by the tetrahedra $\Delta$ and its seven mirror images. The union of the regions corresponding to the tetrahedra formed with all the possible unordered triplets of edges converging in $v$
is denoted by $D(v)$, that is $D(v)\coloneqq \cup_{ v(\Delta)=v }
D(\Delta) $. Taking this into account and assuming that the vertex $v$ has
valence $n$, then the number of possible unordered triplets is
$E(v)=n(n-1)(n-2)/6$, and the relevant contribution to the integral will
be, 
\begin{align}
\sum\limits_{v} \frac{1}{E(v)} \sum\limits_{v(\Delta)=v} \sum\limits_{\Delta' \in D(\Delta)} \int_{\Delta'}.
\end{align}
 Where we already took into account the fact that the Hamiltonian acts only on the vertices of the graph. The first sum is over all vertices. The second in the tetrahedra
formed by all the possible triplets of edges that arrive at the
vertex. The third sum is in the tetrahedron $\Delta$ and its mirror
images, which produces a factor eight. Then, using the triangulation adapted to the graph of the spin
network $\Gamma$ we mentioned (we call it $T(\Gamma)$) we obtain the
quantum expression,
\begin{align}
 \hat{H}^E_T (N) = \sum\limits_{v(\Delta)=v} N_v \frac{8}{E(v)} \sum\limits_{v(\Delta)=v} \hat{H}^E_\Delta = \sum\limits_{v\in V(\gamma)} N_v \hat{H}^E_v .
\end{align}

For the reduced theory the form of the Euclidean parts of the
Hamiltonian are just particular cases of the expression we just
discussed, but we would like to have their expressions in terms of the
reduced variables. The starting point is the classical expression,
\begin{align}
h_E&=\int dxdy \frac{N}{\sqrt{\e}}\left[(\a^{i}_{b,a} - \a^{i}_{a,b} + \epsilon_{ilm}\a^{l}_{a} \a^{m}_{b})\epsilon_{ijk}\e_{j}^{a} \e_{k}^{b} +2\delta^j_3 \delta_b^\phi\left(\a^i_a \e_i^a\e_j^b - \a^i_a\e_i^b \e_j^a\right)\right] \nonumber \\
\label{h_E_axi}
&=2 \int dxdy {N}  \left[\varepsilon^{abc} f_{ab}^i \left\lbrace \a_c^i , V \right\rbrace + 2 \delta_3^j \delta_b^\phi \a_a^i \varepsilon_{ijk} \varepsilon^{a b c} \left\lbrace \a_c^k , V \right\rbrace \right].
\end{align}
It should be noted that the volume that appears here is that of the
complete theory, which can be written in terms of the reduced
variables as $V=\int d^3x \sqrt{{\rm det}({\bf e})}$.

We now observe that the key term
\begin{align}
\label{key_term}
\varepsilon^{abc} f_{ab}^i \left\lbrace \a_c^i , V \right\rbrace
\end{align}
can be divided into two sums,
\begin{itemize}
 \item $a=\alpha,b=\beta,c=\phi$
 \item $a=\alpha,b=\phi,c=\beta ~~~ (\text{and} ~ a=\phi,b=\alpha,c=\beta) $.
 \end{itemize}
 The first sum works as in the full theory except that the term with
 the Poisson bracket will have a point holonomy.
  As in the full theory, we now proceed to discretize \eqref{h_E_axi} in terms of holonomies. Let us first partition the two-surface $\Sigma$ in elementary triangles $\Delta$. Analogously to the full theory, in each triangle we choose one of its vertices and call it $v(\Delta)$. Let $s_i(\Delta)$ with $i=1,2$ be the edges that arrive at $v$ and $\alpha_{ij}=s_i(\Delta)\circ a_{ij}\circ s_j^{-1}(\Delta) $, with $a_{ij}$ the edge that links the endpoints of $s_i$ and $s_j$ which do not coincide with $v(\Delta)$. Consider now the quantity
 \begin{align}
\label{h_triangulacion_axi_sigma}
 h_T^{E,\Sigma} (N) = \sum\limits_{\Delta \in T} h_\Delta^{E,\Sigma} (N),
\end{align}
where we have defined
\begin{align}
\label{h_triangulacion_2_axi_sigma}
 h_\Delta^{E,\Sigma} (N) \coloneqq -\frac{1}{2\pi} N_v \varepsilon^{ijk} \operatorname{tr} \left( h_{\alpha_{ij}(\Delta)} h_{s_k (\Delta)} \left\lbrace h^{-1}_{s_k (\Delta)} , V \right\rbrace \right).
\end{align}
being $s_k$ an edge perpendicular to $\Sigma$ such that one of its endpoints is $v(\Delta)$.
 The presence of the factor of $2\pi$ will become clear shortly . It can be readily seen that when the triangle $\Delta$ shrinks to its base point, the previous quantity tends to
\begin{align}
\label{h_E_axi_discretized}
 2 \frac{\mu}{2\pi} \int_\Delta N \left( \varepsilon^{\alpha\beta\phi} f^i_{\alpha\beta}\{\a_\phi^i , V \} \right)= \frac{1}{2\pi}\int_{\Delta \times \left[ 0,\mu \right]} 2  N \left( \varepsilon^{\alpha\beta\phi} f^i_{\alpha\beta}\{\a_\phi^i , V \} \right)
\end{align}
where we have renamed the indices to facilitate comparison with \eqref{key_term}. This integral is performed in the volume $\Delta\times \left[0,\mu\right]$ instead of a tetrahedron. Recall that when obtaining \eqref{h_E_axi}, one has to perform an integral in the coordinate corresponding to the Killing vector field \cite{kerr1}, which simply yields a factor of $2\pi$ since none of the quantities depend on such variable, leaving us with a two dimensional integral on a surface perpendicular to the k.v.f. Since we partitioned the two surface $\Sigma$ in the triangles $\Delta$, and these surfaces can be \lq\lq carried\rq\rq around the whole space by moving them along the orbits of the Killing vector field (whose orbits are closed spacelike curves), it follows that we can partition our three dimensional space by elemental tori obtained from carrying the elemental triangles $\Delta$ around the orbits of $\partial_\phi$ of length $2\pi$. Partitioning those tori in regions of the form $\Delta\times[0,\mu_i]$ such that $\sum\limits_i \mu_i = 2\pi$ and summing for all $\mu_i$ in \eqref{h_E_axi_discretized} cancels the $2\pi$ factor in the denominator and is the equivalent to performing the integral in $\phi$ that leaves us with the two dimensional integral in \eqref{h_E_axi}. After doing this and taking the sum in \eqref{h_E_axi_discretized} for all $\Delta$ we obtain,
\begin{align}
 \int_{\Sigma} 2  N \left( \varepsilon^{\alpha\beta\phi} f^i_{\alpha\beta}\{\a_\phi^i , V \} \right).
\end{align}

We conclude therefore that \eqref{h_triangulacion_2_axi_sigma} is the
discretized or ``holonomized''  version of the $a=\alpha,b=\beta,c=\phi$ part of the first term in \eqref{h_E_axi}.
\\[30pt]
\begin{figure}
{\includegraphics[width=12cm]{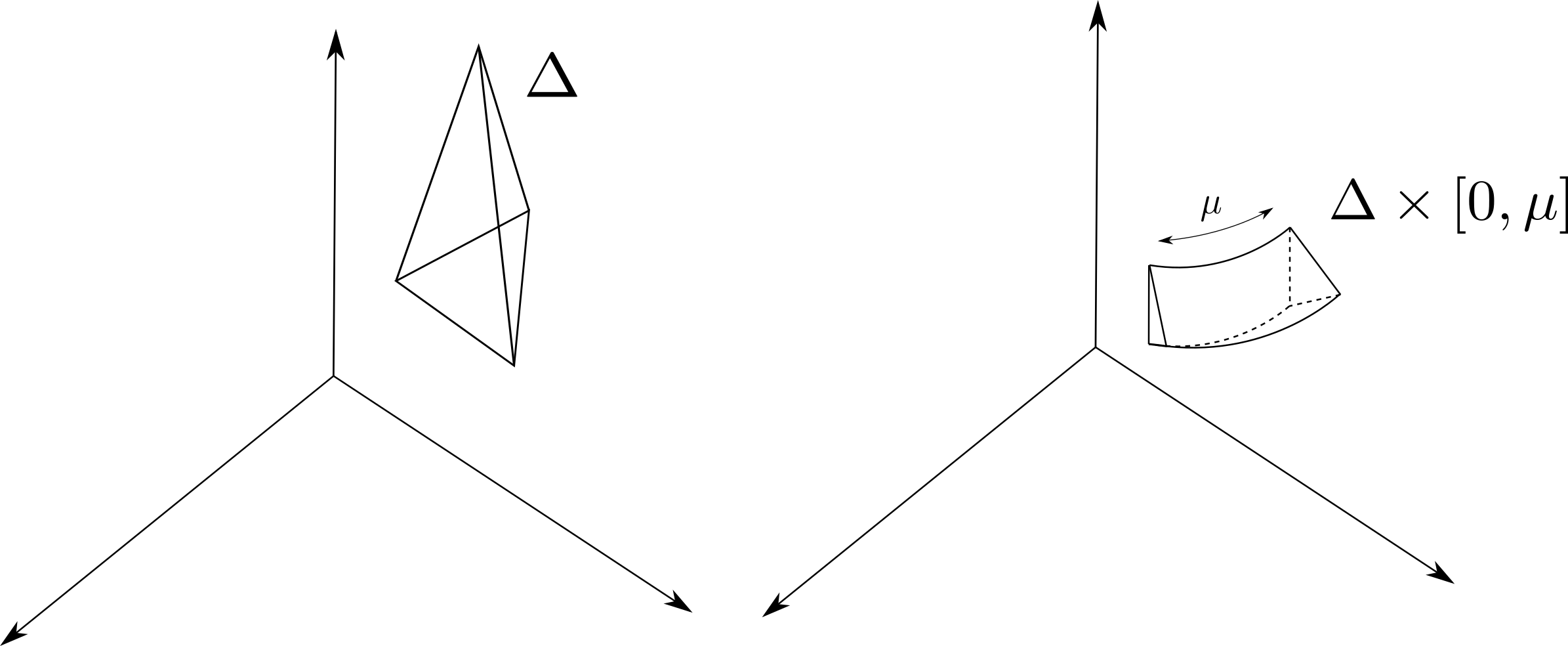}} \\
\caption{In the full theory, one partitions the space into elementary tetrahedra (left), while in the axisymmetric theory we can partition it into elementary triangular tori (right).}
\end{figure}

A discretization for the rest of the euclidean Hamiltonian constraint, namely
\begin{align}
\label{h_E_axi_2}
 h^{E,\phi} &\coloneqq 2\int dxdy {N}  \left[\left(\varepsilon^{\alpha\phi\beta} f_{\alpha\phi}^i \left\lbrace \a_\phi^i , V \right\rbrace + (\alpha \leftrightarrow \phi) \right) + 2 \delta_3^j \delta_b^\phi \a_a^i \varepsilon_{ijk} \varepsilon^{a b c} \left\lbrace \a_c^k , V \right\rbrace \right],
\end{align}
can be obtained in a similar fashion. 
Let us first consider the following product of holonomies:

\begin{align*}
 h^{-1}_{\mu v}(x) h^{-1}_{\varepsilon u}(x+\varepsilon u) h_{\mu v} (x+\varepsilon u) h_{\varepsilon u} (x)
\end{align*}
with $u$ a unit vector in $\Sigma$ and $v$ also unit in the Killing
direction. $x$ is an arbitrary point in $\Sigma$ and dependence in the parentheses denotes the starting point of the holonomy. 
Up to now wherever it says $\a_a^i$ we could have substituted the variable that
transforms properly as a connection \cite{kerr1},
$\tilde{\a}_a^i \coloneqq \a_a^i + \delta_a^\phi \delta^i_3$.  It
makes no difference as it appears inside a Poisson bracket and the
extra term does not contribute. Similarly, the components of $f_{ab}$ considered do not involve the extra term. From now on, however,
it will be better to consider holonomies built explicitly with $\tilde{\a}_a^i$.
Both variables are canonically conjugate to $\e^a_i$. 

Taking now the limit $\varepsilon\to 0$ and $\mu\to 0$ 
\begin{align*}
 h^{-1}_{\mu v}(x) h^{-1}_{\varepsilon u}(x+\varepsilon u) h_{\mu v} (x+\varepsilon u) h_{\varepsilon u} (x) \approx & \left( 1 - \mu \tilde{\a}_a(x) v^a \right)\left( 1 - \varepsilon \tilde{\a}_a(x+\varepsilon u) u^a \right) \\
 &\left( 1 + \mu \tilde{\a}_a(x+\varepsilon u) v^a \right) \left( 1 + \varepsilon \tilde{\a}_a(x) u^a \right) \\
 = & \left( 1 - \mu \tilde{\a}_v(x)  \right)\left( 1 - \varepsilon \tilde{\a}_u(x+\varepsilon u) \right) \\
 &\left( 1 + \mu \tilde{\a}_v(x+\varepsilon u) \right) \left( 1 + \varepsilon \tilde{\a}_u(x) \right).
\end{align*}
Where to keep things brief we defined $\tilde{\a}_a \coloneqq
\tilde{\a}_a^i \tau_i$. Keeping terms up to order $\varepsilon \mu$ (and dropping the dependence in the starting point):
\begin{align}
 h^{-1}_{\mu v} h^{-1}_{\varepsilon u} h_{\mu v} h_{\varepsilon u} \approx & 1+\varepsilon\mu \left( \partial_u \tilde{\a}_v^i + \tilde{\a}_v^j \tilde{\a}^k_u \varepsilon_{jki} \right) \tau_i \nonumber \\
 & 1 + \varepsilon \mu \left( \partial_u {\a}_v^i + \left( {\a}_v^j + \delta^j_3 \right) \a_u^k \varepsilon_{jki} \right) \tau_i.
\end{align}
Now consider the term
\begin{align}
 h^{-1}_{\varepsilon_2 w} \left\lbrace h_{\varepsilon_2 w} , V \right\rbrace \approx \varepsilon_2 \left\lbrace \a_w^i , V \right\rbrace \tau_i,
\end{align}
with $w$ a unit vector in the two-surface, and the limit $\varepsilon_2\to 0$ was taken. 
Using these last two results (and substituting $v\to \phi$, and $u,w\to
\alpha,\beta$) we have that,

\begin{align*}
 \varepsilon^{\alpha\phi\beta} \operatorname{tr}\left(h^{-1}_{\mu\phi} h^{-1}_{\varepsilon \alpha} h_{\mu\phi} h_{\varepsilon\alpha} h^{-1}_{\varepsilon_2 \beta} \left\lbrace h_{\varepsilon_2 \beta} , V \right\rbrace \right\rbrace \approx \varepsilon \varepsilon_2 \mu ~ \varepsilon^{\alpha\phi\beta} \left( f_{\alpha\phi}^i \left\lbrace \a_\beta^i , V \right\rbrace + \varepsilon_{3ji} \a_\alpha^j \left\lbrace \a_\beta^i , V \right\rbrace \right),
\end{align*}
that is,
\begin{align*}
\label{haxi_infinitesimal_angular}
 \varepsilon^{\alpha\phi\beta} \operatorname{tr}\left(h^{-1}_{\mu\phi} h^{-1}_{\varepsilon \alpha} h_{\mu\phi} h_{\varepsilon\alpha} h^{-1}_{\varepsilon_2 \beta} \left\lbrace h_{\varepsilon_2 \beta} , V \right\rbrace \right\rbrace \rightarrow \frac{1}{2} \int_{\Delta\times [0,\mu]} ~ \varepsilon^{\alpha\phi\beta} \left( f_{\alpha\phi}^i \left\lbrace \a_\beta^i , V \right\rbrace + \varepsilon_{3ji} \a_\alpha^j \left\lbrace \a_\beta^i , V \right\rbrace \right).
\end{align*}
Where $\Delta$ is a triangle with one of its vertices in $x$ and two of its sides $ \varepsilon u $ and $ \varepsilon_2 w $.
It is straightforward to check that by interchanging $\alpha$ and $\phi$ we obtain
\begin{align}
 &\frac{1}{2} \int_{\Delta\times [0,\mu]} ~ \varepsilon^{\phi\alpha\beta} \left( f_{\phi\alpha}^i \left\lbrace \a_\beta^i , V \right\rbrace - \varepsilon_{3ji} \a_\alpha^j \left\lbrace \a_\beta^i , V \right\rbrace \right)\nonumber \\
 =& \frac{1}{2} \int_{\Delta\times [0,\mu]} ~ \varepsilon^{\phi\alpha\beta} f_{\phi\alpha}^i \left\lbrace \a_\beta^i , V \right\rbrace + \varepsilon^{\alpha\phi\beta} \varepsilon_{3ji} \a_\alpha^j \left\lbrace \a_\beta^i , V \right\rbrace.
\end{align}
Putting these two together we obtain the term corresponding to
$a=\alpha (\phi), b=\phi (\alpha), c=\beta$ in \eqref{key_term} plus
the last term in \eqref{h_E_axi}.

We are now able to find the Euclidean Hamiltonian operator. Let us
partition the two-surface $\Sigma$ in elementary triangles $\Delta$ as
before. For each triangle we pick one of its vertices and name it
$\Delta(v)$. Let the edges $s_\alpha$ and $s_\beta$ be the two sides
that start at $v$, and $s_\phi$ an edge starting at $v$ and
perpendicular to $\Sigma$.  With the results obtained so far, it is
straightforward to see that the following operator results from the discretization of \eqref{h_E_axi},
 \begin{align}
 \hat{H}_{T}^{E}(N) :=\sum_{\Delta \in T} \hat{H}_{\Delta}^{E}(N) = \sum\limits_{\Delta\in T} \left( \hat{h}_\Delta^\alpha (N) + \hat{h}^\phi_\Delta (N) \right)
\end{align}
with the terms in the sum,
\begin{align}
\label{h_E_alpha_delta}
 \hat{h}^{\alpha}_{\Delta} (N) &:=-2 \frac{N(v(\Delta))}{2 i \ell_{p}^{2}} \epsilon^{\alpha \beta \phi} \operatorname{tr}\left(h_{\alpha_{\alpha \beta}(\Delta)} h_{s_{\phi}(\Delta)}\left[h_{s_{\phi}(\Delta)}^{-1}, \hat{V}\right]\right), \\
 \label{h_E_phi_delta}
 \hat{h}^\phi_\Delta &:=-2 \frac{N(v(\Delta))}{2 i \ell_{p}^{2}} \epsilon^{\alpha \phi \beta} \operatorname{tr}\left( h^{-1}_{s_\phi} h^{-1}_{s_\alpha} h_{s_\phi} h_{s_\alpha} h^{-1}_{s_\beta} \left[ h_{s_\beta} , \hat{V} \right] \right) + \left( \alpha \leftrightarrow \phi \right).
\end{align}
Again this operator only acts on the vertices of the graph $\gamma$ of the
state it is acting on. We conclude this section by (roughly) noting
how do \eqref{h_E_alpha_delta} and \eqref{h_E_phi_delta} act on the
simplest non trivial vertex possible. Let $v\in \gamma$ be a vertex
such that the edges $e_\alpha \supset s_\alpha$ and $e_\beta \supset
s_\beta$ are both outgoing at it, while the edge $e_\phi \supset
s_\phi$ (which is transverse to $\Sigma$) starts and ends at $v$. It
is easy to see that $\hat{V}$ annihilates such a vertex so this
selects the term in the commutator in both \eqref{h_E_alpha_delta} and
\eqref{h_E_phi_delta} in which it acts after the holonomy. We will see
in more detail in a future work that the action of the holonomy which
acts before the volume and its inverse after it can be \lq\lq factored
out\rq\rq; basically, after the first holonomy and the volume have
both acted, the second holonomy gets contracted with the first one
(its inverse) giving as a result a Kronecker delta which simply gives
us the trace of the product of rest of the holonomies. We are
therefore left with the action of the holonomies along the loops
$\alpha_{\alpha\beta}$ and $ s^{-1}_\phi \circ s^{-1}_\alpha \circ
s_\phi \circ s_\alpha  $. Their action on the graph $\gamma$ can be
represented pictorially, as shown in figure 3.

\begin{figure}[h]
{\includegraphics[width=12cm]{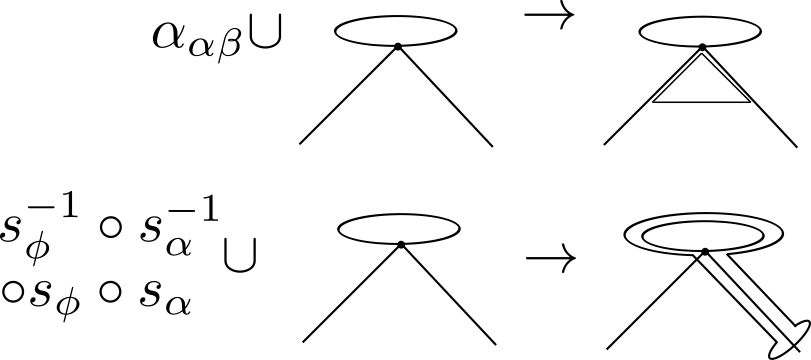}} 
\caption{Action of the Hamiltonian on the graph $\gamma$ when applied
  to a simple vertex. Both terms add a loop to the graph. Although the
  lines should intersect, they have been separated for clarity. In the
diagram, the straight lines live in $\Sigma$ and the ovals represent
circles in the $\phi$ direction of the Killing vector field.}
\end{figure}

\section{Lorentzian part of the Hamiltonian cons\-traint}

In the full theory the Lorentzian part of the Hamiltonian constraint
is given by \cite{thiemann},

\begin{align}
\label{H_L}
 H_{L}(N)&=2 \int d^{3} x N\left(1+\beta^{2}\right) \frac{1}{\sqrt{E}} \epsilon_{i j k} \epsilon_{i l m} E_{j}^{a} E_{k}^{b} K_{a}^{l} K_{b}^{m} \nonumber
 \\
 &= 4 \int d^{3} x N\left(1+\beta^{2}\right) \varepsilon^{a b c} \varepsilon_{i j k}\left\{A_{a}^{j}, K\right\}\left\{A_{b}^{k}, K\right\}\left\{A_{c}^{i}, V\right\},
\end{align}
with $K\coloneqq E^a_i K_a^i$. This quantity can in turn be written as
\begin{align}
\label{K}
 K=-\left\{V, \int_{\Sigma} d^{3} x {\cal H}_{E}(x)\right\}=-\left\{V, H_{E}(1)\right\},
\end{align}
where $\cal H$ is the Hamiltonian density.
An analogous expression holds for the axisymmetric case:
\begin{align}
\label{h_L}
h_{L}(N) &=2 \int d^{3} x N\left(1+\beta^{2}\right) \frac{1}{\sqrt{\e}} \epsilon_{i j k} \epsilon_{i l m} \e_{j}^{a} \e_{k}^{b} k_{a}^{l} k_{b}^{m} \nonumber
\\ 
&=4 \int d^{3} x N\left(1+\beta^{2}\right) \varepsilon^{a b c} \varepsilon_{i j k}\left\{ \a_{a}^{j}, k\right\}\left\{ \a_{b}^{k}, k\right\}\left\{ \a_{c}^{i}, V\right\} .
\end{align}
Again we could have used $\tilde{\a}_a^i$, or ${\a}_a^i$.
It is also easy to show the following identity holds:
\begin{align}
\label{k_identity}
 k=-\left\{V, \int_{\Sigma} d^{3} x \mathcal{ h}_{E}(x)\right\},
\end{align}
with $\mathcal{h}$ the Hamiltonian density.
To see this, we first note that
\begin{align}
\label{k}
 k\coloneqq k_a^i \e^a_i.
\end{align}
Now recall that $\beta k_a^i= \a_a^i - \gamma_a^i$, with $\gamma_a^i$
the spin connection,
\begin{align}
 \gamma_{a}^{i}=\frac{1}{2} \epsilon_{i j k} \mathbf{e}_{j}^{b}\left(\mathbf{e}_{a, b}^{k}-\mathbf{e}_{b, a}^{k}+\mathbf{e}_{k}^{c} \mathbf{e}_{a}^{l} \mathbf{e}_{c, b}^{l}+\mathbf{e}_{a}^{k} \mathbf{e}_{c}^{l} \mathbf{e}_{l, b}^{c}\right)-\delta_{3}^{i} \delta_{a}^{\phi}:=\tilde{\gamma}_{a}^{i}-\delta_{3}^{i} \delta_{a}^{\phi},
\end{align}
using this result we define
$\beta k_a^i=\a_a^i-\tilde{\gamma}_a^i+\delta^i_a \delta^\phi_a \coloneqq
\beta \tilde{k}_a^i + \delta^i_a \delta^\phi_a $ so that \eqref{k} can be
written as,
\begin{align}
\beta k=\beta k_{a}^{i} \mathbf{e}_{i}^{a}=\beta \tilde{k}_{a}^{i} \mathbf{e}_{i}^{a}+\mathbf{e}_{3}^{\phi}.
\end{align}
Now if we write the Euclidean Hamiltonian as $h_E = h_\alpha  +h_\phi $ so that,
\begin{align}
\label{v_h_E_axi}
 -\left\{V, \int_{\sigma} d^{2} x \mathcal{h}_{E}(x)\right\}=-\left\{V, \int_{\sigma} d^{2} x \mathcal{h}_{\alpha}(x)\right\}-\left\{V, \int_{\sigma} d^{2} x \mathcal{h}_{\phi}(x)\right\}.
\end{align}
It is evident that the first term corresponds to $\tilde{k}\coloneqq
\tilde{k}_a^i \e^a_i$ since it is completely analogous to
\eqref{K}. Now for the second term let us write $h_\phi$ explicitly:
\begin{align}
 h_{\phi}(N)=\frac{1}{8 G} \int \mathrm{d}^{2} x \frac{N}{\sqrt{\mathrm{e}}} 2 \delta_{3}^{j} \delta_{b}^{\phi}\left(\mathbf{a}_{a}^{i} \mathbf{e}_{i}^{a} \mathbf{e}_{j}^{b}-\mathbf{a}_{a}^{i} \mathbf{e}_{i}^{b} \mathbf{e}_{j}^{a}\right).
\end{align}

From the following identity
\begin{align}
 \frac{1}{\sqrt{\e}} \epsilon^{i j k} \e_{j}^{a} \e_{k}^{b}=2 \epsilon^{a b c}\left\{ \a_{c}^{i}, V\right\},
\end{align}
we can obtain
\begin{align}
 \left\{V, \mathbf{a}_{a}^{i}\right\}=- \frac{1}{2} \sqrt{\mathbf{e}} \mathbf{e}_{a}^{i},
\end{align}
and using this, the second term in \eqref{v_h_E_axi} is:
\begin{align}
 -\left\{V, h_{\phi}[1]\right\}=\frac{1}{2} \mathbf{e}_{a}^{i} \delta_{3}^{j} \delta_{b}^{\phi}\left(\mathbf{e}_{i}^{a} \mathbf{e}_{j}^{b}-\mathbf{e}_{i}^{b} \mathbf{e}_{j}^{a}\right)=\frac{1}{2}\left(3 e_{3}^{\phi}-\mathbf{e}_{i}^{\phi} \delta_{3}^{i}\right)=\mathbf{e}_{3}^{\phi}
\end{align}
so \eqref{k_identity} holds as claimed. 

The steps for finding a discretization for the Lorentzian portion of
the Hamiltonian constraint are identical to those followed in the Euclidean case so we will not repeat them here. We partition the two-surface $\Sigma$ as before in elementary triangles $\Delta$ and define,
\begin{align}
\label{h_L_Delta}
 h_{\Delta}^{L}(N)=\frac{4}{2} N(v(\Delta)) \left(1+\beta^{2}\right) \varepsilon^{i j k} \operatorname{tr}\left(h_{s_{i}}^{-1}\left\{h_{s_{i}}, k\right\} h_{s_{j}}^{-1}\left\{h_{s_{j}}, k\right\} h_{s_{k}}^{-1}\left\{h_{s_{k}}, V\right\}\right),
\end{align}
where two of the edges $s_i,s_j,s_k$ are sides of the triangle joining
at $v(\Delta)$ while the other one is perpendicular to $\Sigma$. It is
straightforward to check that when $v(\Delta)$ shrinks to its base
point $v$, this previous expression tends to $ 4 (1+\beta^2) \int_{\Delta\times
  [0,\mu]} N \operatorname{tr}\left( \left\lbrace {\bf a},k \right\rbrace
  \left\lbrace {\bf a},k \right\rbrace \left\lbrace {\bf a},V \right\rbrace
\right) $, therefore a discretization for \eqref{h_L} is given by:
\begin{align}
 h^L(N)=\sum\limits_{\Delta} h^L_\Delta (N).
\end{align}
The quantization of this expression is obtained by adapting the triangulation to a graph $\gamma$ as was done for the Euclidean case, promoting functions to operators and replacing the Poisson brackets by commutators. The resulting operator has the following action on a function $\Psi$ cylindrical with respect to $\gamma$:
\begin{eqnarray}
\hat{h}^L(N) \Psi&=&-\frac{16}{2\left(i \ell_{p}^{2}\right)^{3}} \sum_{v \in V(\gamma)} N_{v} \frac{1}{E(v)} \sum_{v(\Delta)=v} \epsilon^{i j k} \times\nonumber \\ &&\times \operatorname{tr}\left(h_{s_{i}(\Delta)}\left[h_{s_{i}(\Delta)}^{-1}, k_{v}\right] h_{s_{j}(\Delta)}\left[h_{s_{j}(\Delta)}^{-1}, k_{v}\right] h_{s_{k}}(\Delta)\left[h_{s_{k}(\Delta)}^{-1}, V_{v}\right]\right) \Psi \nonumber \\ &=:& \sum_{v \in V(\gamma)} \frac{N_{v}}{E(v)} \sum_{v(\Delta)=v} \hat{h}^L_{\Delta} \Psi =: \sum_{v \in V(\gamma)} \frac{N_{v}}{E(v)} \hat{h}^L_{v} \Psi,
\end{eqnarray}
where $k_v$ and $V_v$ indicates they are evaluated at the vertex $v$ and 
we have already used that the operator only acts on the vertices
of $\gamma$ (the argument used in the Euclidean case can be repeated
here) and also the fact that the commutators with holonomies along
edges adjacent to $v$ are non vanishing only when $k$ and $V$ are
evaluated on the same $v$ \cite{thiemann}. The vertex multiplicity in
this case is given by $E(v)=n(v)(n(v)-1) / 2$.  

\section{Requirements for a triangulation adapted to a graph}
\subsection{Cylindrical consistency}
All the operators we have defined above depend on a choice of
partition. In the chosen prescription, the partition $T$ is adapted to
a graph $\gamma$, so if $\gamma \subset \gamma'$, $T(\gamma)$ and
$T(\gamma')$ will be different from each other. We should require that
$T$ is consistently defined (possibly up to a diffeomorphism), that
is, if $f$ is a cylindrical function of
$\gamma$, then $\hat{h}_{T(\gamma)} f$ and $ \hat{h}_{T(\gamma')} f$
should be diffeomorphic to each other. This is easily accomplished by
introducing the following operator $\hat{p}_e$ associated with an edge
$e$ whose action on a function $f$ cylindrical with respect to a graph
$\gamma$ is to annihilate it if $\gamma$ and $e$ do not intersect in a
finite segment having the same endpoint as $e$ and to leave
it invariant otherwise. We call this operator the edge projector. With
this definition we can construct a projector associated with a
triangle (in $\Sigma$) from the partition chosen:
$\hat{p}_{\Delta}=\hat{p}_{s_1(\Delta)}
\hat{p}_{s_2(\Delta)}\hat{p}_{s_{\phi}(\Delta)}$ ($s_1(\Delta)$ and
$s_2(\Delta)$ are edges from the partition which converge at
$v(\Delta)$, while $s_\phi(\Delta)$ is an edge in the direction of the
Killing vector field outgoing at $v(\Delta)$) and with it a vertex
projector: $\hat{E}(v)= \sum\limits_{v(\Delta)=v} \hat{p}_\Delta $. We
can now define the following self-consistent operator (up to a
diffeomorphism):
\begin{align}
\label{h_E_consistent}
\hat{h}^E_\gamma \coloneqq \sum\limits_{v\in V(\gamma)} 4 N_v \sum\limits_{v(\Delta)=v}\hat{h}^E_\Delta \frac{\hat{p}_\Delta}{E(v)} .
\end{align}
Given two graphs $\gamma$ and $\gamma'$ such that $\gamma \subset
\gamma'$ and a function $f$ cylindrical with respect to $\gamma'$, it
is clear from \eqref{h_E_consistent} that the terms in
$\hat{h}^E_\gamma$ corresponding to triangles formed with at least an
edge not present in $\gamma$, or such that the edge perpendicular to
the 2-surface based at the corresponding vertex is not present in
$\gamma$ will vanish when applied to $f$. Also, the factor $E(v)$
reduces to the correct value on $f$. That $\hat{h}^E_{\gamma'} f$ and
$\hat{h}^E_{\gamma} f$ are related by a diffeomorphism follows from
the manifestly diffeomorphism invariant prescription for
loop-assigments (see \cite{thiemann} for details) when acting on a
function $f$ cylindrical in $\gamma$.

The Lorentzian Hamiltonian involves only $\hat{k}$ and holonomies with respect to the edges of the original graph $\gamma$. It follows that if $\hat{h}^L$ is cylindrically consistently defined up to a diffeomorphism provided we introduce as before the projectors $\hat{p}_\Delta$:
\begin{align}
 \hat{h}^L_v \coloneqq \sum\limits_{v(\Delta)=v} \hat{h}^L_\Delta \frac{\hat{p}_\Delta}{E(v)}
\end{align}
\subsection{Diffeomorphism Covariance}
When looking for solutions to the Euclidean Hamiltonian constraint, we will consider distributions $\Psi$ in the space of cylindrical functions such that,
\begin{align}
\label{dif_cov_1}
 \Psi [\hat{h}^E (N) f] = 0 ~ \forall N ~ \forall f \in \operatorname{Cyl}^\infty (\mathcal{A}/\mathcal{G}).
\end{align}
 We want this quantity to depend only on the diffeomorphism invariant
 pro\-per\-ties of the triangulation assignment, which can be
 accomplished if we can make the quantum Hamiltonian be diffeomorphism
 covariant. If $f$ is cylindrical with respect to a graph $\gamma$,
 then $\hat{h}^E f$ will be a linear combination of cylindrical
 functions with respect to the graph $T(\gamma)$, now if $f'$ is
 cylindrical with respect to a graph $\gamma'$ related to $\gamma$ by
 a diffeomorphism $\gamma'=\phi(\gamma)$, the quantity $\hat{h}^E f'$ will consist of cylindrical functions with respect to the graph $T(\phi(\gamma))$. We will require that $T(\gamma)$ and $T(\phi(\gamma))$ be diffeomorphic to each other. Let us begin by observing that \eqref{dif_cov_1} can be written as
\begin{align}
\label{dif_cov_2}
 \Psi [\sum\limits_{v} \hat{h}^E_v f]= \sum\limits_{v} \Psi [ \hat{h}^E_v f] = 0 ~ \forall \gamma ~ \forall f\in\operatorname{Cyl}^\infty (\mathcal{A}/\mathcal{G}) 
\end{align}
so that the requirement of diffeomorphism covariance can be formulated
in terms of the Hamiltonian evaluated at each vertex $v$. It is easy to
see that for every vertex, each triangle $\Delta$ (with
$v(\Delta)=v$) of the corresponding partition can be treated
separately. Moreover, each term $\hat{h}^E_\Delta$ is the sum of
$\hat{h}^{E,\alpha}_\Delta$ and $ \hat{h}^{E,\phi}_\Delta $ which can
also be treated separately. The projectors are already covariantly
defined so we will focus on the rest of the operator. 

Let's start by making $ \hat{h}^{E,\alpha} $ covariantly defined. This
case is analogous to that of the full theory \cite{thiemann}. If two
graphs $\gamma$ and $\gamma'$ are related by a diffeomorphism $\phi$,
then the triangles $\Delta(\phi(\gamma)), \phi(\Delta(\gamma))$ will
in general be different. More specifically, if $s_1$ and $s_2$ are
segments of the edges $e_1 \subset \gamma$ and $e_2 \subset\gamma$
both outgoing at $v$, then the graph $T(\gamma)$ is formed by adding
to $\gamma$ an arc $a$ joins the endpoints of $s_1$ and $s_2$ which do
not coincide with $v$. When acting with $\phi$ on $T(\gamma)$, the
edges $s_i, e_i ~ (i=1,2)$ and $a_{12}$ will be mapped to $s'_i, e'_i
~ (i=1,2)$ and $a'_{12}$ respectively. Now consider the graph
$\phi(\gamma)$, to obtain $T(\phi(\gamma))$ we add an arc $
\tilde{a}_{12} $ joining two segments $\tilde{s}_i \subset e'_i, ~
i=1,2$, which are not necessarily equal to $s'_i, ~ i=1,2$, therefore,
the arcs $a'_{12}$ and $\tilde{a}_{12}$ will in general differ. Recall
however that the arcs depend only on the topology of the vertex, and
the graphs $\gamma, ~ \gamma'$ are diffeomorphic to each other. What
we need is a diffeomorphism $\phi'$ that leaves $\gamma'$ invariant
while $\phi' (\Delta(\gamma')) = \phi(\Delta(\gamma))$. Such $\phi'$
clearly exists: It is the diffeomorphism that leaves the image of the
graph invariant while moving its points within itself it can put the
arc $\tilde{a}_{12}$ into any diffeomorphic shape. By using the
$\phi'$ corresponding to each $\Delta$ in \eqref{dif_cov_2}, we can
make $\hat{h}^{E,\alpha}$ diffeomorphism covariant. All of the above
can be summed up in figure 4.

\begin{figure}[h]
{\includegraphics[width=12cm]{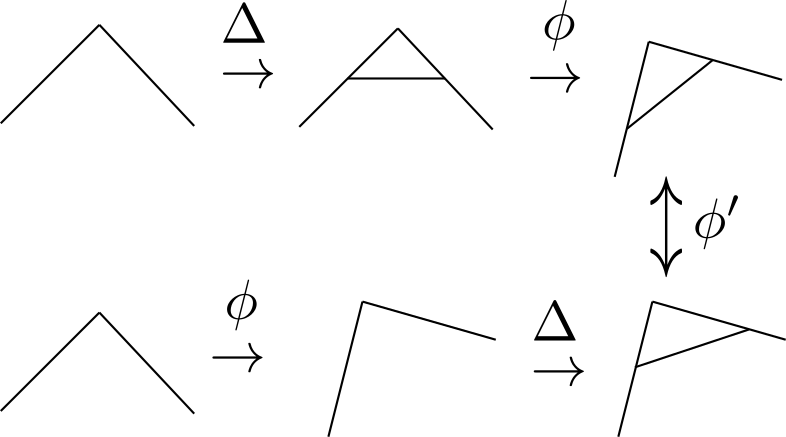}} \\
\caption{Simple graphical explanation of the paragraph above. The perpendicular edge has been omitted for simplicity. The diffeomorphism $\phi^{\prime}$ doesn't change the original graph but moves its points within itself, thus being able to put the triangles into any diffeomorphic shape.}
\end{figure}

The procedure we can follow for the term $\hat{h}^{E,\phi}$ is very
similar, we begin by noting that it involves holonomy terms out of
which only one is taken along an edge not belonging to the original
graph $\gamma$. Applying $\hat{h}^{E,\phi}_\Delta$ to the same
$f_\gamma$ as before, we obtain a linear combination of functions
cylindrical with respect to a graph $T(\gamma)$ which will differ from
the original graph by a segment $s_\phi$, transverse to the two
surface. (There is a subtlety that will be discussed later, but it
doesn't affect the present discussion: basically, $T(\gamma)$ differs
from $\gamma$ by a closed loop \lq\lq with a nose\rq\rq that
intersects with it everywhere except at the segment $s_\phi$). This
new segment intersects $\gamma$ at a point, which is the endpoint of a
segment $s_\alpha\subset e$, where $e \subset \gamma$ is an edge
contained in the two surface and outgoing at $v(\Delta)$. Acting with
a diffeomorphism $\phi$ on the graph $T(\gamma)$ yields
$\phi(T(\gamma))$ while $s, s_\phi$ and $e$ where mapped to $s',
s'_\phi$ and $e'$ respectively. Now consider the function
$f_{\phi(\gamma)}$, after acting with $h^{E,\phi}_{\phi(\Delta)}$ we
obtain functions cylindrical with respect to $\Delta(\phi(\gamma))$,
which differs from $ \phi(\Delta(\gamma)) $ by an edge $s''_\phi$,
which is transverse to the two surface and intersects $\phi(\gamma)$
at the endpoint of $s'' \subset e'$ that doesn't coincide with the
vertex. Neither $s''$ nor $s''_\phi$ are necessarily equal to $s'$ and
$s'_\phi$ so $ T(\phi(\gamma)) $ and $ \phi(T(\gamma)) $ will in
general differ. Similar to the previous case, we would like to have a
diffeomorphism such that $\phi'(\phi(\gamma))=\phi(\gamma)$ and
$\phi'(\Delta(\phi(\gamma)))=\phi(\Delta(\gamma))$. The required
diffeomorphism again exists and its action on $\phi(\gamma)$ is to
leave the graph image invariant but move its points such that the
intersection point $s'' \cap s''_\phi$ can be moved within $e'$ until
it coincides with $s' \cap s'_\phi$. Pictorially, this would
correspond, in the diagram on the lower right of figure 3, to shorten
the edges that connect the two ovals. This is all we need to make the
term $\hat{h}^{E,\phi}$ diffeomorphism covariant. 

Finally, since we have managed to make the entire Euclidean
Hamiltonian diffeomorphism covariant, the Lorentzian Hamiltonian's
covariance follows tri\-via\-lly since it is defined in terms of
$\hat{k}$ and holonomies along edges of the original graph, so it will
be covariantly defined if and only if $\hat{h}^{E}$ is.

\subsection{Anomaly freeness}

The considerations in the previous sections have allowed us to redefine the Hamiltonian operator in a cylindrically consistent, diffeomorphism covariant way. There is still some freedom in our prescription for a triangulation, the operator depends on the diffeomorphism class of the triangulation assignment. We should check now if our Hamiltonian is anomaly free, that is, if 
\begin{align}
\label{anomaly_freeness}
 \Psi \left[ \left[\hat{h}(N),\hat{h}(M)\right] f \right] = 0.
\end{align}
For any diffeomorphism invariant distribution $\Psi$, cylindrical function $f$ and lapses $N$ and $M$ \cite{thiemann}. Since 
\begin{align}
\label{anomaly_freeness2}
\hat{h}(N)=\hat{h}^E(N) + \hat{h}^L(N),
\end{align} 
and moreover,
\begin{align}
\label{anomaly_freeness3}
\hat{h}^E(N)=\hat{h}^{E,\alpha}(N)+\hat{h}^{E,\phi}(N),
\end{align}
trying to verify anomaly freeness directly in \eqref{anomaly_freeness} would be (although straightforward) very tedious, even more so than in the case of the full theory. Out of all the terms that would appear in \eqref{anomaly_freeness} if we substituted \eqref{anomaly_freeness2} and \eqref{anomaly_freeness3}, we will focus just on one of them, namely,
\begin{align}
\label{anomaly_freeness4}
 \Psi \left[ \left[\hat{h^{E,\alpha}}(N),\hat{h}^{E,\alpha}(M)\right] f \right],  
\end{align}
and show that it equals zero for every distribution $\Psi$ and every cylindrical function $f$. This particular term is perhaps the easiest to analyze since it resembles
\begin{align}
\label{anomaly_freeness4}
 \Psi \left[ \left[\hat{H^{E}}(N),\hat{H}^{E}(M)\right] f \right] ,
\end{align}
where $\hat{H}^E$ is the Euclidean Hamiltonian operator from the full
theory \cite{thiemann}. Moreover, after proving that the operator
$\hat{h}^{E,\alpha}$ is anomaly free, a moment of reflexion can
convince one that every single term appearing in the decomposition of
\eqref{anomaly_freeness} mentioned above can be shown to vanish
separately, thus proving that the entire Hamiltonian operator is
anomaly free. To see this, we will use the following key observation:
Every term in the linear combination forming the Hamiltonian operator,
when acting on a function cylindrical with respect to a certain graph,
produces a linear combination of functions cylindrical with respect to
a graph with additional vertices; however, none of the terms of the
Hamiltonian acts on these new vertices. 

Let us begin by recalling the definition of $\hat{h}^{E,\alpha}$. Given a graph $\gamma$ and a triangulation $T(\gamma)$:
\begin{align}
 \hat{h}^{E,\alpha}=\sum\limits_{v=V(\gamma)} \frac{N_v}{E(v,\gamma)} \sum\limits_{v(\Delta)=v}\hat{h}^{E,\alpha}_{\Delta} \hat{p}_{\Delta}.
\end{align}

Now, if $f$ is a cylindrical function based on $\gamma$, and we have
that $v= v(\Delta) \neq v(\Delta')=v'$, then
$\hat{p}_{\Delta'} \hat{h}^{E,\alpha}_\Delta f =
\hat{h}^{E,\alpha}_\Delta f $ since the only possible removed edges
in $\gamma$ are in $v$. If follows that
$\hat{E}(v') \hat{h}^{E,\alpha}_\Delta f = E(v',\gamma)
\hat{h}^{E,\alpha}_\Delta f$, that is, all the vertices adjacent to
$v'$ in the original graph $\gamma$ are still present, so their
multiplicity remains the same after acting with
$\hat{h}^{E,\alpha}_\Delta$. If however $v'=v$, then
$ \hat{h}^{E,\alpha}_\Delta f $ is a linear combination of cylindrical
functions whose graphs may have missing edges in $v$, in those cases
$\hat{E}(v') \hat{h}^{E,\alpha}_\Delta f \neq E(v',\gamma)
\hat{h}^{E,\alpha}_\Delta f$. 

Next let us compute,
\begin{align}
 & \hat{h}^{E,\alpha}(N) \hat{h}^{E,\alpha}(N) f = \sum_{v \in V(\gamma)} \frac{N_{v}}{E(v, \gamma)} \sum_{v(\Delta(\gamma))=v} \hat{h}^{E,\alpha}(N) \hat{h}_{\Delta(\gamma)}^{E,\alpha} f \nonumber
 \\
 &=\sum_{v \in V(\gamma)} \frac{N_{v}}{E(v, \gamma)} \sum_{v(\Delta(\gamma))=v} \sum_{v^{\prime} \in V(\gamma \cup \Delta(\gamma))} M_v \sum_{v(\Delta(\gamma \cup \Delta(\gamma)))=v^{\prime}} \hat{h}_{\Delta(\gamma \cup \Delta(\gamma))}^{E,\alpha} \frac{1}{\hat{E}\left(v^{\prime}\right)} \hat{h}_{\Delta(\gamma)}^{E,\alpha} f.
\end{align}
If we compute the same quantity with the roles of $N$ and $M$ exchanged and then subtract:
 \begin{align}
&\left[\hat{h}^{E,\alpha}(N), \hat{h}^{E,\alpha}(N)\right] f \nonumber
\\ 
=&\sum_{v, v^{\prime} \in V(\gamma)}\left(M_{v^{\prime}} N_{v}-M_{v} N_{v^{\prime}}\right) \frac{1}{E(v, \gamma)}  \sum_{v(\Delta(\gamma))=v, v(\Delta(\gamma \cup \Delta(\gamma)))=v^{\prime}} \hat{h}_{\Delta(\gamma \cup \Delta(\gamma))=v^{\prime}}^{E,\alpha} \frac{1}{\hat{E}(v')} \hat{h}_{\Delta(\gamma)}^{E,\alpha} f \nonumber
\\ 
=&\sum_{v, v^{\prime} \in V(\gamma), v<v^{\prime}} \frac{M_{v^{\prime}} N_{v}-M_{v} N_{v^{\prime}}}{E(v) E\left(v^{\prime}\right)} \times \nonumber \\ 
&\times \quad \sum_{v\left(\Delta\left(\gamma \cup \Delta^{\prime}(\gamma)\right)\right)=v, v\left(\Delta^{\prime}(\gamma)\right)=v\left(\Delta^{\prime}(\gamma \cup \Delta(\gamma))\right)=v^{\prime}}\left[\hat{h}_{\Delta^{\prime}(\gamma \cup \Delta(\gamma))}^{E,\alpha} \hat{h}_{\Delta(\gamma)}^{E,\alpha}-\hat{h}_{\Delta\left(\gamma \cup \Delta^{\prime}(\gamma)\right)}^{E,\alpha} \hat{h}_{\Delta^{\prime}(\gamma)}^{E,\alpha}\right] f
\end{align}
where we have used the fact that the second Hamiltonian term doesn't
act on the vertices added to the graph by the first. We were able to
substitute $\hat{E}({v'})$ by $ E(v') \coloneqq E(v',\gamma) $, since
the only possible case in which $\hat{E}(v') \hat{h}^{E,\alpha} f \neq
E(v',\gamma) \hat{h}^{E,\alpha}_{\Delta(\gamma)} f$ is when $v'=v$
(there may be some missing edges in the graph of some of the terms in
the linear combination $\hat{h}^{E,\alpha}_{\Delta(\gamma)} f$), but
those terms do not contribute thanks to the antisymmetrized lapses
product. 

To verify anomaly freeness, it will be sufficient to show that,
\begin{align}
 \label{anomaly_freeness5}
\Psi\left[\left(\hat{h}_{\Delta^{\prime}(\gamma \cup \Delta(\gamma))}^{E,\alpha} \hat{h}_{\Delta(\gamma)}^{E,\alpha}-\hat{h}_{\Delta\left(\gamma \cup \Delta^{\prime}(\gamma)\right)}^{E,\alpha} \hat{h}_{\Delta^{\prime}(\gamma)}^{E,\alpha}\right) f\right]=0,
\end{align}
holds for each choice of $v(\Delta(\gamma))=v\left(\Delta\left(\gamma \cup \Delta^{\prime}(\gamma)\right)\right)=v, v\left(\Delta^{\prime}(\gamma)\right)=v\left(\Delta^{\prime}(\gamma \cup \Delta(\gamma))\right)=v^{\prime}$. Now, since $\hat{h}_{\Delta^{\prime}(\gamma \cup \Delta(\gamma))}$ doesn't really act on the vertices added by $ \hat{h}^{E,\alpha}_{\Delta(\gamma)} $ and besides $v\neq v'$, we can find a diffeomorphism $\phi^{\prime}$ such that $\phi^{\prime} \left( \Delta' (\gamma \cup \Delta(\gamma)) \right) = \Delta^{\prime}(\gamma) $ and $\phi'(\Delta(\gamma))= \Delta(\gamma) $. Similarly, we can find another diffeomorphism $\phi$ such that $\phi(\Delta(\gamma \cup \Delta' (\gamma)))=\Delta(\gamma)$ while $\Delta' (\gamma)$ is left invariant. It follows that we can write or previous claim \eqref{anomaly_freeness5} as
\begin{align}
\label{anomaly_freeness6}
\Psi\left[\left(\hat{U}(\phi') \hat{h}_{\Delta^{\prime}(\gamma)}^{E,\alpha} \hat{h}_{\Delta(\gamma)}^{E,\alpha}- \hat{U}(\phi) \hat{h}_{\Delta\left(\gamma\right)}^{E,\alpha} \hat{h}_{\Delta^{\prime}(\gamma)}^{E,\alpha}\right) f\right]=0.
\end{align}

The state $\Psi$ is diffeomorphism invariant so both $\hat{U}(\phi')$ and $ \hat{U}(\phi) $ can be removed and since $\hat{h}^{E,\alpha}_{(\Delta(\gamma))}$ and $ \hat{h}^{E,\alpha}_{(\Delta'(\gamma))} $ act on different vertices they commute. We have thus verified \eqref{anomaly_freeness5}. \\
The reasoning followed to analyze the rest of the terms in \eqref{anomaly_freeness} involving $\hat{h}^{E,\alpha}$ and $\hat{h}^L$ is essentially the same. Repeating the same steps, we will arrive at a term analogous to \eqref{anomaly_freeness5}, in which the operator to the left will correspond to a triangle with vertex $v'$ in the triangulation of a graph $\gamma'$ bigger than the original $\gamma$. The operator to the right acts on a different vertex $v$ and the triangulation is adapted to the original graph. Removing those diffeomorphisms and taking into account the fact that the operators in each term act on different vertices and therefore commute, we obtain de desired result. Following this same line of reasoning for all the remaining terms, we are able to verify our initial claim \eqref{anomaly_freeness}.

\section{Conclusion}

We have constructed a quantum theory for axisymmetric space-times
based on a symmetry reduction of the Ashtekar formulation of gravity 
we introduced in a previous paper. We defined the kinematical states,
and the actions of the basic operators, area and volume. With them we
proceeded to construct the quantum version of the Euclidean and
Lorentzian parts of the Hamiltonian constraint. We showed their
cylindrical consistency and anomaly-freeness of the constraint
algebra. In a further paper we will present the explicit evaluation of
the Hamiltonian constraint and discuss its space of solutions. 

The parallelism with the full theory is remarkable. This has
advantages and disadvantages. On the one hand, it does not appear that
working in one dimension less due to the symmetry reduces the
complexities of the action of the Hamiltonian. In fact, it requires
treating two different kinds of terms of comparable complexity to
those of the full theory. On the other hand, it appears that the steps
taken to include matter in a anomaly free way in the full theory could
be implemented in this case. It should be noted that in the case of
even one dimension less than the case considered in this paper, 
spherically symmetric reductions, up to today
there is no known way to include matter without modifying the
constraint algebra \cite{bojo}

\section*{Acknowledgment}
We wish to thank Javier Olmedo 
for discussions. This work was supported in part by Grants NSF-PHY-1603630,
NSF-PHY-1903799, funds of the Hearne Institute for Theoretical
Physics, CCT-LSU, fqxi.org, Pedeciba, and CAP-CSIC.


\begin{thebibliography}{99}
\bibitem{kerr1}  R.~Gambini, E.~Mato, J.~Olmedo and J.~Pullin,
  Class.\ Quant.\ Grav.\  {\bf 36}, no. 12, 125009 (2019)
  doi:10.1088/1361-6382/ab1d82
  [arXiv:1812.05403 [gr-qc]].
\bibitem{thiemann} T. Thiemann, ``Modern canonical general
  relativity'', Cambridge University Press, Cambridge, UK (2008).
\bibitem{varadarajan}  M.~Varadarajan,
  Phys.\ Rev.\ D {\bf 100}, no. 6, 066018 (2019)
  doi:10.1103/PhysRevD.100.066018
  [arXiv:1904.02247 [gr-qc]].
\bibitem{lewandowski}
  M.~Assanioussi, J.~Lewandowski and I.~M\"akinen,
  Phys.\ Rev.\ D {\bf 92}, no. 4, 044042 (2015)
  doi:10.1103/PhysRevD.92.044042
  [arXiv:1506.00299 [gr-qc]];
    J.~Lewandowski and H.~Sahlmann,
  Phys.\ Rev.\ D {\bf 91}, no. 4, 044022 (2015)
  doi:10.1103/PhysRevD.91.044022
  [arXiv:1410.5276 [gr-qc]].
\bibitem{wald_book}
  R.M.~Wald, General Relativity,
  Chicago Univ. Press,
  1984
  \bibitem{asle_volume}
    A.~Ashtekar, J.~Lewandowski, 
   Adv.\ Theor.\ Math.\ Phys. 1 (1998)388-429,doi: 10.4310/ATMP.1997.v1.n2.a8
  [gr-qc/9711031].
  \bibitem{aslemamoth}
    A.~Ashtekar, J.~Lewandowski, D.~Marolf, J.~Mourao and T.~Thiemann,
  J.\ Math.\ Phys.\  {\bf 36}, 6456 (1995)
  doi:10.1063/1.531252
  [gr-qc/9504018].
  
\bibitem{bojo}
  M.~Bojowald, S.~Brahma and J.~D.~Reyes,
  Phys.\ Rev.\ D {\bf 92}, no. 4, 045043 (2015)
  doi:10.1103/PhysRevD.92.045043
  [arXiv:1507.00329 [gr-qc]].
\end{thebibliography}
\end{document}